\documentstyle[prl,aps,preprint]{revtex}

\let\ssection=\section
\renewcommand{\section}{\setcounter{equation}{0}\ssection}

\begin{document}
\draft 
\title{ON THE RELATION BETWEEN QUADRATIC AND LINEAR CURVATURE
  LAGRANGIANS IN POINCAR\'E GAUGE GRAVITY\footnote{In: {\it New Ideas in
    the Theory of Fundamental Interactions.} Proceedings of the Second
    German--Polish Symposium, Zakopane 1995. H-D.\ Doebner, M.\ 
    Paw{\l}owski, R.\ R{\c{a}}czka (eds.) to be published.}} \author{Yuri N.
  Obukhov\footnote{On leave from: Department of Theoretical Physics,
    Moscow State University, 117234 Moscow, Russia.} and Friedrich W.
  Hehl} \address{Institute for Theoretical Physics, University of
  Cologne,\\ D--50923 K\"oln, Germany}
\maketitle

\begin{abstract}
  We discuss the choice of the Lagrangian in the Poincar\'e gauge
  theory of gravity. Drawing analogies to earlier de Sitter gauge
  models, we point out the possibility of deriving the Einstein-Cartan
  Lagrangian {\it without} cosmological term from a {\it modified}
  quadratic curvature invariant of {\it topological} type.
\end{abstract}
\bigskip
\pacs{PACS no.: 04.50.+h; 04.20.Cv}

\section{Introduction}

One of the main achievements of the gauge approach to gravity (see
\cite{is}$^{\!-\,}$\cite{erice95} and references therein) lies in a 
better understanding of the deep relations between the symmetry groups
of spacetime and the nature of the sources of the gravitational field.
At the same time a satisfactory {\it kinematical} picture of gauge
gravity emerges which specifies metric, coframe, and connection as the
fundamental gravitational field variables. In contrast, the {\it
dynamical} aspect of the gravitational gauge theory is far less
developed.  In general, the choice of a dynamical scheme, i.e.\ of the
gravitational field Lagrangian, ranges from the simplest
Einstein-Cartan model with the Hilbert type Lagrangian (linear in
curvature) to the 15-parameter theory with Lagrangian quadratic in
torsion and curvature, or even to non-polynomial models.

Some progress was achieved in gauge theories based on the de Sitter
group~\cite{mm}$^{\!-\,}$\cite{gh} which is, in a sense, the closest
semi-simple ``relative'' of the (non-semi-simple) Poincar\'e group.
The main idea behind the derivation of the gravitational field
Lagrangian was to consider it as emerging, via a certain spontaneous
breakdown symmetry mechanism, from a unique invariant, the
Chern-Pontryagin or the Euler invariant, e.g., which have the meaning
of topological charges. Recently this approach has been reanalyzed in
\cite{nieto}.

In this paper we report on an attempt to exploit the analogy with the
de Sitter gauge approach. In a Riemann-Cartan spacetime, we construct
a gravitational Lagrangian by starting from a topological invariant
quadratic in curvature, deform it suitably, and arrive, apart from an
exact form, at an Einstein-Cartan Lagrangian (linear in curvature).
Whereas in the traditional approach of (\ref{e1}) the emergence of a
cosmological term cannot be prevented, our new method, see our main
result (\ref{p1}), yields a pure Einstein-Cartan Lagrangian {\it without}
cosmological term.

\section{Two four-dimensional topological invariants}

The spacetime which we consider obeys a Riemann-Cartan geometry with
{\it orthonormal} coframe $\vartheta^\alpha$, a metric $g= o_{\alpha\beta}
\,\vartheta^\alpha\otimes\vartheta^\beta$, and a Lorentz connection
$\Gamma^{\alpha\beta}=-\Gamma^{\beta\alpha}=\Gamma_i{}^{\alpha\beta}dx^i$.
Here the anholonomic frame indices are denoted by $\alpha,\beta,\dots=
\hat{0},\hat{1},\hat{2},\hat{3}$, the holonomic coordinate indices by
$i,j,\dots = 0,1,2,3$, and $o_{\alpha\beta}=diag\,(-1,+1,+1,+1)$ is the
local Minkowski metric (with the help of which we raise and lower
Greek indices).

As it is well known, in four dimensions there are two topological
invariants which are constructed from the (in general, Riemann-Cartan)
curvature two-form $R_\alpha{}^\beta=d\Gamma_\alpha{}^\beta-
\Gamma_\alpha{}^\gamma\wedge\Gamma_\gamma{}^\beta$.  These are the
{\it Euler} invariant defined by the four-form
\begin{equation}
E :=R_{\alpha\beta}\wedge{R}^{\star\alpha\beta}= 
{1\over 2}\eta^{\alpha\beta\mu\nu}R_{\alpha\beta}\wedge R_{\mu\nu},\label{eu}
\end{equation}
and the {\it Chern--Pontryagin} invariant described by the four-form
\begin{equation}
  P:=-R_\alpha{}^\beta\wedge R_\beta{}^\alpha =
  R_{\alpha\beta}\wedge{R}^{\alpha\beta}.\label{pon}
\end{equation}
Both forms (\ref{eu}) and (\ref{pon}) are functionals of the Lorentz
connection $\Gamma^{\alpha\beta}$ and of the local metric
$o_{\alpha\beta}$. The $\eta^{\alpha\beta\mu\nu}$ is the Levi-Civita
tensor, and no any other geometrical variables are involved. The right
star $^\star$ denotes the so-called Lie dual with respect to the Lie
algebra indices.

The Gauss-Bonnet theorem states that an integral of (\ref{eu}), with a
proper normalization constant, over a compact manifold without a
boundary describes its Euler characteristics (the alternating sum of
the Betti numbers which count the simplexes in an arbitrary
triangulation of the manifold). As for the integral of (\ref{pon}),
also introducing proper normalization, this represents the familiar
``instanton'' number specialized to the gravitational gauge case.

\section{Ordinary and twisted deformations of the curvature}

Due to the peculiar properties of the Lie algebra of the de Sitter
group, a new object appears within the framework of de Sitter gauge
gravity part of the generalized $SO(1,4)$ or $SO(2,3)$ curvature,
a two-form
\begin{equation}
  {\Omega}_{\alpha\beta}:=R_{\alpha\beta}- {1\over
    \ell^2}\,\vartheta_{\alpha\beta} \,,\qquad{\rm with}\qquad
  \vartheta_{\alpha\beta}:=\vartheta_{\alpha}\wedge\vartheta_\beta
  \,.\label{omega}
\end{equation}
We may call it a deformation of the original curvature form by a
specific contribution constructed from the translational gauge
potentials, namely the coframe one-forms $\vartheta^{\alpha}$. The
constant $\ell$ with the dimension of length provides the correct
dimension. If we recall that the curvature of a Riemann-Cartan
spacetime can be decomposed into six irreducible pieces
$^{(N)}R_{\alpha\beta}$, with $N=1,\dots,6$, see \cite{new}, then we
find that $\vartheta_{\alpha\beta}$ is proportional to
the sixth pieces $^{(6)}R_{\alpha\beta}$, the curvature scalar, that
is, in (\ref{omega}) we subtracted a certain constant {\it scalar}
curvature piece from the total curvature.

Similarly to (\ref{omega}), by means of the Hodge star, we can define
another deformation
\begin{equation}
  {\cal R}_{\alpha\beta}:=R_{\alpha\beta}- {1\over \ell^2}\,
  \eta_{\alpha\beta}= R_{\alpha\beta}-{1\over \ell^2}\,^\ast\!\left(
\vartheta_\alpha\wedge\vartheta_\beta\right) .\label{rr}
\end{equation}
We may call this a twisted translational deformation, since
$\eta_{\alpha\beta}$ has the opposite parity behavior compared to
$R_{\alpha\beta}$. In fact, the term $\eta_{\alpha\beta}$ is
proportional to a constant pseudoscalar piece
$^{(3)}R_{\alpha\beta}$ of the curvature or, in components, to
$R_{[\gamma\delta\alpha\beta]}$. In other words, in (\ref{rr}) a
constant {\it pseudoscalar} curvature piece is subtracted out. Note that
$^{(3)}R_{\alpha\beta}$ vanishes together with the torsion since, by
means of the first Bianchi identity, 
\begin{equation}
  D T^\alpha =R_\beta{}^\alpha\wedge\vartheta^\beta\qquad{\rm or}\qquad
  DT^\alpha\wedge \vartheta_\alpha=R_{\alpha\beta}\wedge
  \vartheta^\alpha\wedge\vartheta^\beta=\,^{(3)}R_{\alpha\beta}\wedge
  \vartheta^\alpha\wedge\vartheta^\beta\,.\label{b1a}
\end{equation} 
Hence, in a Riemannian spacetime, $^{(3)}R_{\alpha\beta}$ vanishes
identically.

Using also the irreducible decomposition of the torsion into three pieces
$^{(M)}T^\alpha$, with $M=1,2,3$, the last equation can be rewritten as
\begin{eqnarray}
  ^{(3)}R_{\alpha\beta}\wedge\vartheta^\alpha\wedge\vartheta^\beta=
  d\left(\vartheta_\alpha\wedge T^\alpha\right)\!\!&-&\!\!T_\alpha\wedge
  T^\alpha\nonumber\\= d\left(\vartheta_\alpha\wedge\,^{(3)}
    T^\alpha\right)\!\!&-&\!\!^{(1)}T_\alpha\wedge \,^{(1)}T^\alpha
-2\,^{(2)}T_\alpha\wedge \,^{(3)}T^\alpha\,.\label{b1b}
\end{eqnarray}
For a proof of this equation see \cite{new} Eq.(B.2.19). 

Before we consider some Lagrangians in the next section, we develop
some algebra for the quadratic expressions of
$\vartheta_{\alpha\beta}$ and $\eta_{\alpha\beta}$. We find:
\begin{equation}
  \vartheta_{\alpha\beta}\wedge\vartheta^{\alpha\beta}=
  -\left(\vartheta_\alpha\wedge\vartheta^\alpha\right)\wedge\left(
    \vartheta_\beta\wedge\vartheta^\beta\right)=0\,.\label{th2}
\end{equation}
Moreover, for any two-form $\Phi$, we have $^{\ast\ast}\Phi=-\Phi$.
Consequently, we find
\begin{equation}
  \vartheta_{\alpha\beta}\wedge\vartheta^{\alpha\beta}=
  -\left(^{\ast\ast}\vartheta_{\alpha\beta}\right)\wedge
  \vartheta^{\alpha\beta}=-\,^\ast\vartheta_{\alpha\beta}\wedge\,^\ast
  \vartheta^{\alpha\beta}=-\eta_{\alpha\beta}\wedge\eta^{\alpha\beta}
\label{et2}
\end{equation}
or
\begin{equation}
\eta_{\alpha\beta}\wedge\eta^{\alpha\beta}=0
\,.\label{et2'}
\end{equation}
The mixed term can be expanded as follows:
\begin{equation}
  \vartheta_{\alpha\beta}\wedge\eta^{\alpha\beta}=\frac{1}{2}\,
  \eta^{\alpha\beta\gamma\delta}\vartheta_{\alpha\beta}\wedge
  \vartheta_{\gamma\delta}=12\,\eta\,.\label{thet}
\end{equation}
Here $\eta$ is, as usually, the volume four-form. If we transform the
Hodge star $^\ast$ into the Lie star $^\star$, we have
\begin{equation}
  \eta^{\alpha\beta}=\vartheta^{\star\alpha\beta}=
  \,^\ast\vartheta^{\alpha\beta} \,.\label{star}
\end{equation}
Eventually, we take the Lie star of the last equation:
\begin{equation}
  \eta^{\star\alpha\beta}=-\vartheta^{\alpha\beta}\,.\label{star'}
\end{equation}
 
\section{Deformed topological invariants \- and the Einstein-Cartan Lagrangian}

Let us calculate the Euler and Pontryagin four-forms with the
curvature replaced by the deformed curvature. We denote the Lagrangian
of the Einstein-Cartan theory by
\begin{equation}
  L_{\rm EC}:=-{1\over 2\ell^2}\,\eta_{\alpha\beta}\wedge
  R^{\alpha\beta}\,. \label{ec}
\end{equation}
Later we will meet similar Lagrangians with $\eta_{\alpha\beta}$ substituted 
by $\vartheta_{\alpha\beta}$. We do the corresponding algebra first:
\begin{eqnarray}
  \vartheta_{\alpha\beta}\wedge R^{\alpha\beta}&=&\frac{1}{2}\,R^{\gamma
    \delta\alpha\beta}\,\vartheta_{\gamma\delta}\wedge\vartheta_{\alpha\beta}
\nonumber \\
  &=&\frac{1}{2}\,R_{[\alpha\beta\gamma\delta]}\,\vartheta^\alpha\wedge
\vartheta^\beta\wedge\vartheta^\gamma\wedge\vartheta^\delta=\,
^{(3)}R_{\alpha\beta}\wedge\vartheta^\alpha\wedge\vartheta^\beta
\,,\label{ranti}\\
\eta_{\alpha\beta}\wedge R^{\alpha\beta}&=&-2\ell^2L_{\rm EC}
\,,\label{ec*}\\
  \vartheta_{\alpha\beta}\wedge R^{\star\alpha\beta}&=&\vartheta^{\star
    \alpha\beta}\wedge R_{\alpha\beta}= \eta_{\alpha\beta}\wedge
  R^{\alpha\beta}=-2\ell^2L_{\rm EC} \,,\label{ec'}\\
  \eta_{\alpha\beta}\wedge R^{\star\alpha\beta}&=&
  \eta^{\star\alpha\beta}\wedge R_{\alpha\beta}= -
  \vartheta^{\alpha\beta}\wedge R_{\alpha\beta}=- \,
  ^{(3)}R_{\alpha\beta}\wedge\vartheta^\alpha\wedge\vartheta^\beta
  \,.\label{ranti'}
\end{eqnarray}
In the formulas (\ref{ranti}) and (\ref{ranti'}) it is of course
possible to substitute the first Bianchi identity (\ref{b1b}) in order
to splitt off a boundary term, if desireable.

For the deformations (\ref{omega}) and (\ref{rr}) one finds,
respectively, the following generalized Euler forms:
\begin{eqnarray}
V_{\rm Eu}&=& \Omega_{\alpha\beta}\wedge\Omega^{\star\alpha\beta}=
E + 4L_{\rm EC} + {12\over\ell^4}\,\eta,\label{e1}\\
  V'_{\rm Eu}&=& {\cal R}_{\alpha\beta}\wedge{\cal
    R}^{\star\alpha\beta}= E + {2\over\ell^2}\,
  ^{(3)}R_{\alpha\beta}\wedge\vartheta^\alpha\wedge\vartheta^\beta
-{12\over\ell^4}\,\eta
\,,\label{e2}\\
  V''_{\rm Eu}&=&{\Omega}_{\alpha\beta}\wedge{\cal R}^{\star\alpha
    \beta}= E+2L_{\rm EC}+\frac{1}{\ell^2}\, ^{(3)}R_{\alpha\beta}
  \wedge \vartheta^\alpha\wedge\vartheta^\beta \,.\label{e3}
\end{eqnarray}
It is interesting to note that the translational Chern--Simons term
$\vartheta_\alpha\wedge T^\alpha$ \cite{tcs}, via (\ref{b1b}), appears
as boundary term in (\ref{e2}) and (\ref{e3}). The other mixed term,
${\cal R}_{\alpha\beta}\wedge {\Omega}^{\star\alpha\beta}$, is the
same as that in (\ref{e3}), since the Lie star can be moved to ${\cal
  R}_{\alpha\beta}$.

Three more generalized topological Lagrangians are defined according to  the
Chern-Pontryagin pattern as follows:
\begin{eqnarray}
  V_{\rm Po}&=&{\cal R}_{\alpha\beta}\wedge{\cal R}^{\alpha\beta}= P +
  4L_{\rm EC},\label{p1} \\ 
 V'_{\rm Po}&=&\Omega_{\alpha\beta}\wedge\Omega^{\alpha\beta}= P -
  \frac{2}{\ell^2}\,^{(3)}R_{\alpha\beta}\wedge\vartheta^\alpha\wedge
  \vartheta^\beta ,\label{p2} \\ 
 V''_{\rm Po}&=&{\Omega}_{\alpha\beta}\wedge{\cal R}^{\alpha\beta}=
  P+2\,L_{\rm EC}-\frac{1}{\ell^2}\,^{(3)}R_{\alpha\beta}\wedge
  \vartheta^\alpha\wedge \vartheta^\beta+\frac{12}{\ell^4}\,\eta .\label{p3}
\end{eqnarray}

As we can see, both deformed curvatures, (\ref{omega}) and (\ref{rr}),
generate the Einstein-Cartan Lagrangian (\ref{ec}) from the
topological type invariants, since the variational derivatives of $E$
and $P$ are identically zero. In the case of $\Omega_{\alpha\beta}$
one should use the Euler type form (\ref{e1}), while for ${\cal
  R}_{\alpha\beta}$ the Chern-Pontryagin type invariant (\ref{p1})
suggests itself. Actually, the case (\ref{e1}) was studied in the work
of MacDowell and Mansouri \cite{mm} 
(see also~\cite{town}$^{\!-\,}$\cite{nieto}).
The problem of this de Sitter gauge approach was a very large
cosmological constant $\sim {1/\ell^4}$ which is generated
simultaneously with the Einstein-Cartan Lagrangian. To the best of our
knowledge, the possibility (\ref{p1}) of using the twisted deformation
of the curvature was not reported in the literature, even if Mielke
\cite{pseudo} had somewhat related thoughts, see his Eq.(9.8). A nice
improvement of the usual de Sitter result is then the {\it absence of
  the cosmological term} in (\ref{p1}). It can certainly happen that
the cosmological constant would reappear due to other physical
mechanisms (through the quantum vacuum corrections, e.g.), but the
huge initial value ${1/\ell^4}$ is avoided.

The inspection of the Lagrangians (\ref{e2}) and (\ref{p2}) shows that
they are trivial from the dynamical point of view. Since, up to a
boundary term, $^{(3)}R_{\alpha\beta}\wedge\vartheta^\alpha\wedge
\vartheta^\beta$ $\sim T_\alpha\wedge T^\alpha$, see (\ref{b1b}), the
vacuum field equations leave the curvature undetermined while the
torsion turns out to be zero. In the non-vacuum case, like in the
Einstein-Cartan theory, the torsion is related to the spin current.
More exactly, it is proportional to the Hodge dual of the spin. A
similar thing happens in the curvature sector where the left hand side 
of the gravitational field equation is then represented not by the
Einstein form but rather by its dual. This theory evidently has no
Newtonian limit and thus appears to be physically irrelevant.

The cases (\ref{e3}) and (\ref{p3}) also induce an Einstein-Cartan
Lagrangian.  However, in both cases additionally a definite parity
violating term \cite{stueck,mukku} emer\-ges, see also \cite{bianchi}
Sec.5.3, a possibility which one has to keep in mind, but presently a
need for such terms is not obvious. The Lagrangian (\ref{p3})
represents the MacDowell-Mansouri Lagrangian with an additional parity
violating admixture, whereas (\ref{e3}) is attached to our new
Lagrangian (\ref{p1}) in an analogous way.

It would be tempting to include the twisted deformation of the
curvature (\ref{rr}) into a generalized gravitational gauge theory
analogously to the way (\ref{omega}) appears in the de Sitter model.
However this seems to be impossible, at least at the present level of
understanding this problem.  A simple argument runs as follows: The de
Sitter group describes the symmetry of a four-dimensional model
spacetime the curvature of which is defined by putting (\ref{omega})
equal zero, i.e. the model is a de Sitter spacetime with
$R^{\alpha\beta}={1\over\ell^2}\vartheta^{\alpha}\wedge
\vartheta^{\beta}$. Calculating the exterior covariant derivative of
this equation, one finds that, {\it in four dimensions}, torsion is
equal to zero, and one is left with the usual Riemannian spacetime of
constant curvature.  Unlike this, the condition
$R^{\alpha\beta}={1\over\ell^2}\eta^{\alpha\beta}$ specifies a
spacetime of constant pseudoscalar curvature, but does not seem to
define any sound spacetime geometry. Again taking the covariant
exterior derivative, one discovers that torsion is absent, and the
remaining equation, which involves the purely Riemannian curvature,
turns out to be inconsistent. Hence it looks as if no fundamental
spacetime existed with a symmetry property which would make it
possible to include the twisted deformation of the curvature in some
sector of a generalized gauge group.

\section{Scalar field}

In general case, the third and sixth irreducible parts of the Riemann-Cartan
curvature read \cite{new}
\begin{equation}
{}^{(3)}R_{\alpha\beta}=-{1\over 12}X\eta_{\alpha\beta}, \quad\qquad
{}^{(6)}R_{\alpha\beta}=-{1\over 12}R\,\vartheta_{\alpha}
\wedge\vartheta_{\beta}, \label{decomp}
\end{equation}
where the curvature pseudoscalar and scalar are defined by
\begin{equation}
X:={\ }^{\ast}(\vartheta_{\alpha}\wedge\vartheta_{\beta}\wedge 
R^{\alpha\beta})\,\qquad{\rm or}\qquad 
R:=e_{\alpha}\rfloor e_{\beta}\rfloor R^{\alpha\beta}\,,\label{scal}
\end{equation}
respectively. This suggests a natural generalization of the
deformations (\ref{omega}) and (\ref{rr}) by introducing a scalar
field $\Phi$ according to
\begin{equation}
{\Omega}^{\Phi}_{\alpha\beta}:=R_{\alpha\beta}- \Phi^{2}
\vartheta_{\alpha\beta}\label{omega1}
\end{equation}
and\begin{equation}
{\cal R}^{\Phi}_{\alpha\beta}:=R_{\alpha\beta}- 
\Phi^{2}\eta_{\alpha\beta}\,.\label{rr1}
\end{equation}
There is no need to introduce a constant factor $\ell^{-2}$ since the
canonical dimension of a scalar field is already $\ell^{-1}$.

The analysis of the arising gravitational Lagrangians is
straightforward, since it is only necessary to replace in the formulas
(\ref{e1})-(\ref{p3}) everywhere ${1/\ell}$ by $\Phi$. In the
{\it absence of matter} one notices immediately that all these models
possess the Weyl conformal symmetry
\begin{eqnarray}
  \vartheta^{\alpha}&\longrightarrow &
  e^{\lambda}\vartheta^{\alpha}\,,\qquad {\rm hence}\qquad
  g=o_{\alpha\beta}\,\vartheta^{\alpha}\otimes\vartheta^{\beta}
  \longrightarrow e^{2\lambda}g\,,\label{conf1}\\ \Phi&
  \longrightarrow & e^{-\lambda}\Phi\,.\label{conf2}
\end{eqnarray}
With the suitable choice of the conformal parameter function $\lambda
$, it is always possible to eliminate the scalar field completely by
picking the gauge $\Phi={1/\ell}$. Thus the Lagrangians
(\ref{e1})-(\ref{p3}) provide a Poincar\'e gauge gravity analog of
the Riemannian gravity sector in the model of Paw\l owski and
R\c{a}czka \cite{pr,PW}. Introducing the matter sector in such a way so
that the conformal symmetry (\ref{conf1})-(\ref{conf2}) is preserved,
it is possible to arrive at the Riemann-Cartan generalization of the
Paw\l owski and R\c{a}czka model. If, however, the coupling with matter
is introduced without respecting (\ref{conf1})-(\ref{conf2}), we
discover the non-Riemannian generalizations of the scalar--tensor
theroies of gravity.

\section{Discussion}

One may notice that both deformations (\ref{omega}) and (\ref{rr}) are
particular cases of the general deformation
\begin{equation}
R^{\alpha\beta}-{A\over\ell^2}\,\vartheta^{\alpha}\wedge\vartheta^{\beta}
-{B\over\ell^2}\,\eta^{\alpha\beta},\label{gen}
\end{equation}
with arbitrary constants $A$ and $B$. Then a straightforward
calculation demonstrates that both topological forms, of the Euler
type as well as of the Chern-Pontryagin type, generate the same
gravitational Lagrangian which includes the true topological
invariant, the Einstein-Cartan term modified by the square of torsion
(``twisted'' Einstein-Cartan) term, plus the cosmological constant
$\sim {1/\ell^4}$. One can verify that the Lagrangian $A\,L_{\rm EC} +
B\,T^{\alpha}\wedge T_{\alpha}$ has the same physical contents as the
usual Einstein-Cartan model, provided $A\neq\pm iB$.  This confirms
earlier observations made within the framework of the self-dual
two-form approaches to gravity theory. In accordance with the
discussion above, in the most general case $A$ and $B$ may be
(nonconstant) scalar fields, thus yielding either conformal invariant
or scalar-tensor versions of the ``twisted'' Einstein-Cartan model.

The general deformation (\ref{gen}) is again closely related to the de
Sitter symmetry group, and hence one could probably obtain this from a
kind of a twisted de Sitter gauge theory. However, the gravitational
Lagrangian should inevitably contain a large classical cosmological
constant with all the known difficulties of physical interpretation of
such a model.


\bigskip
{\bf Acknowledgements} 
\bigskip

We are grateful to Frank Gronwald for helpful discussions. One of the
authors (FWH) would like to thank Ryszard R{\c{a}}czka for the
invitation to the Polish-German Theoretical Physics Meeting in
Zgorzelisko (near Zakopane) in September 1995 where a part of this
paper was drafted. The work of YNO was supported by the grant He
528/17-1 of the Deutsche Forschungsgemeinschaft, Bonn.

\end{document}